# On Enforcing Function Diagram Commutativity And Anti-Commutativity Constraints in *MatBase*

Christian Mancas[1], Diana Christina Mancas[2]

*Ovidius University at Constanta, Romania*

**Corresponding Author:** *Christian Mancas, Bd. Pipera 1/U, Voluntari, 077190, IF, Romania.*

**ABSTRACT**

Presented are algorithms for enforcing function diagram commutativity and anti-commutativity database constraints, using the database software application constraint-driven design and development methodology, in the realm of the (Elementary) Mathematical Data Model ((E)MDM). *MatBase*, an intelligent data and knowledge management system prototype mainly based on the (E)MDM, uses these algorithms to automatically generate corresponding code in both its versions (i.e., the MS Access and the .NET and SQL Server ones). Of course, any software developer may also use these algorithms manually. The paper also discusses the code generated to enforce two such constraints from a Geography database.

## Introduction

The main task of database (db) software applications (apps) must always be enforcing the business rules governing their subuniverses of interest that cannot be enforced by the Database Management Systems (DBMS) managing the corresponding dbs. Not enforcing even a single one of them might result in storing unplausible data in the dbs and this is what eventually happens all the time in such cases.

Relational DBMSes (RDBMS), i.e., those based on the Relational Data Model (RDM, [1, 2, 3]), provide only five types of (relational) constraints for enforcing business rules, namely: domain (range), not null, unique keys, foreign keys (referential integrity), and tuple (check). Besides pure SQL [2, 3, 4] (which is a purely logic programming language augmented with only the UNION algebraic operator) the most powerful of them, e.g., IBM Db/2 [5], Oracle Database [6], MS SQL Server [7], etc., also provide extended SQL versions (SQL PL [8], PL/SQL [9], T-SQL [10], respectively), i.e., pure SQL augmented with procedural programming language constructs (e.g., variable declarations, if-then-else, while, for statements, event-driven triggers, etc.), which might be used to enforce non-relational constraints.

However, such an architectural approach has several drawbacks, ranging from portability concerns (as there is no standard for extending SQL), to efficiency, and, especially, to answering to the following architectural question: why sending unplausible data to RDBMS engines when the business logic of the apps may reject any such data? This is why we too always advocate the dual architectural choice: enforcing the non-relational constraints by using the event-driven methods of high-level programming languages embedding (pure) SQL: these languages (e.g., the .NET family ones [11], MS VBA [12], PHP [13], JavaScript [14]) include at least one statement that has a text-type parameter through which SQL statements are passed to RDBMS engines.

Moreover, for small RDBMSes (e.g., MS Access [12], Oracle MySQL [15], etc.) this alternative is the only one possible, as they do not provide extended SQL.

The (Elementary) Mathematical Data Model ((E)MDM, [16, 17]) currently includes 76 types of constraints,

grouped in the following four categories: set, mapping, object, and system; in the mapping category there are five subcategories: general, function product, homogeneous binary function product (hbfp), self-map, and function diagram (fd). This paper presents the algorithms used by *MatBase* [16, 18, 19], an intelligent data and knowledge base management system prototype mainly based on (E)MDM, the Entity-Relationship (E-R) Data Model [3, 20, 21], and RDM, for enforcing two types of fd constraints: the commutativity and anti-commutativity ones.

The paper is standardly structured: the next section describes the materials and methods used, while section 3 presents and discusses the obtained results; the paper ends with conclusion and the references list. As VBA code is much easier to understand than any of the .NET programming languages, including VBA.Net, we present in section 3 only the results obtained by the MS Access *MatBase* version; however, the algorithms presented in section 2 were also implemented in C# for the MS .NET and SQL Server version of *MatBase*.

**Related Work**

*MatBase* algorithms for detecting and classifying E-R diagram cycles were presented in [22]. To our knowledge, neither the fd commutativity, nor the anti-commutativity db constraint types were studied, except for [23]. Briefly discussed were only cases of local commutativities and anti-commutativities (i.e., composed self-map reflexivities and irreflexivities, respectively) [24, 25], as well as hbfp reflexivities and irreflexivities [26]. Non-relational constraint enforcement is done in *MatBase* according to the apps db constraint-driven design and development paradigm [27].

The most closely related approaches to non-relational constraint enforcement are based on business rules management (BRM) [28, 29] and their corresponding implemented systems (BRMS) and process managers (BPM), like IBM Operational Decision Manager [30], IBM Business Process Manager [31], Red Hat Decision Manager [32], Agiloft Custom Workflow/BPM [33], etc. They are generally based on XML (but also on the Z notation, Business Process Execution Language, Business Process Modeling Notation, Decision Model and Notation, or the Semantics of Business Vocabulary and Business Rules), which is the only other field of endeavor trying to systematically deal with business rules, even if informally, not at the db design level, but at the software application one, and without providing automatic code generation. From this perspective, *MatBase* is also a BRMS, but a formal, automatically code generating one.

## Materials and Methods

A *commutative function diagram constraint* (cfdc) is a function equality of the type $f = f_1 \circ \ldots \circ f_n = g_1 \circ \ldots \circ g_m = g$, where $n$ and $m$ are strictly positive naturals. For example, Figure 7 from [23] presents such a constraint, namely *Continent = Continent ° Range ° Subrange ° Group ° Mountain*, formalizing the rule "any river that springs from a mountain belongs to the same continent as that mountain" (where $n = 1$, $m = 5$). As *Mountain* and even *Group* might take null values, this example is, in fact, a *null-commutative function diagram constraint*.

In the (E)MDM, its case when $n = m = 1$ is also viewed as a hbfp (null-)reflexivity constraint and its enforcement in *MatBase* is provided in [26].

Moreover, the case when $f$ or $g$ is the unity mapping of a set is called in the (E)MDM a *local commutative fd constraint*. For example, Figure 1 from [24] presents such a constraint, namely *State ° StateCapital* = $\mathbf{1}_{STATES}$, formalizing the rule "the state capital of any state must be a city of that state". As such fd constraints are of the type $(\forall x \in D)(f(x) = x)$, where $f$ is a self-map (autofunction), they are of the type self-map reflexivity, whose enforcement is dealt with in [24].

Consequently, in this paper we only deal with the general case of cfdcs, where either $n > 1$ or $m > 1$ and neither $f$ nor $g$ are self-maps. Obviously, changing the values of any of the functions $f_i$, $1 \leq i \leq n$, or $g_j$, $1 \leq j \leq m$, with non-null values might violate such constraints.

To enforce them, *MatBase* first generates in the classes corresponding to the forms built on the db tables $dom(f_n)$ and $dom(g_m)$, for the data row sources of the combo-boxes $f_n$ and $g_m$, SQL statements of the type presented in Figures 1 (when $n > 1$ or $m > 1$) and 2 (when $n = 1$ or $m = 1$), respectively (where $dom(h)$ and $codom(h)$ are the domain and codomains sets of a function $h$, $name(set)$ is the name of a function defined on *set* storing the names of the corresponding objects, and $x$ are the standard (E)MDM names of the object identifiers of any fundamental object set, implemented in the corresponding rdb table by a surrogate primary key; if any of these classes are missing, *MatBase* first creates them).

---

SELECT $f_n$.x, $dom(f_1).name(dom(f_1))$ & " " & … & " " & $dom(f_n).name(dom(f_n))$

　　AS [$name(dom(f_1))$, …, $name(dom(f_n))$], $dom(f_1).f_1$

FROM $dom(f_1)$ RIGHT JOIN ($dom(f_2)$ RIGHT JOIN (… $dom(f_n)$ ON $dom(f_n).f_{n-1} = dom(f_n).x$)

　　… ) ON $dom(f_1).x = dom(f_2).f_2$

ORDER BY $dom(f_1).name(dom(f_1))$ & " " & … & " " & $dom(f_n).name(dom(f_n))$;

**Figure 1.** *SELECT statement generated by MatBase for the row source of $f_n$ when $n > 1$*

> SELECT $codom(g_1).x$, $codom(g_1).name(codom(g_1))$ FROM $codom(g_1)$
> ORDER BY $name(codom(g_1))$;

**Figure 2.** *SELECT statement generated by MatBase for the row source of $g_m$ when m = 1*

Then, *MatBase* creates any class missing from the ones associated to the forms built on the rdb tables $dom(f_1)$, …, $dom(f_n)$, $dom(g_1)$, …, $dom(g_m)$, if any, and, in all these classes, methods of the type $f\_BeforeUpdate$ (for its MS Access version) or $f\_Validating$ (for its MS .NET version), if any. Please recall that these event-driven methods are automatically called by the system whenever the value of $f$ was changed and the user tries to save it (either explicitly, or implicitly, by trying to move the cursor to another control or data line). As *Validating* might be messed up with *Validated* (the equivalent of *AfterUpdate*), we prefer here *BeforeUpdate*.

Then, *MatBase* generates in the *Form_BeforeUpdate* method of the class associated to the common domain of $f$ and $g$ (i.e., the one of $f_n$ and $g_m$) the code shown in Figure 3. Please recall that these event-driven methods are automatically called by the system whenever at least one value of the current data line was changed and the user tries to save it in the corresponding rdb table (either explicitly, or implicitly, by trying to move the cursor to another data line).

> *Method Form_BeforeUpdate(Boolean Cancel)*
> *Cancel = False*
> . . .
> // enforces constraint $f_1 \circ \ldots \circ f_n = g_1 \circ \ldots \circ g_m$
> *if not Cancel and not IsNull($f_n$) and not IsNull($g_m$) then*
>   *if not IsNull($f_n$.Column(2)) and not IsNull($g_m$.Column(2)) then*
>     *if $f_n$.Column(2) ≠ $g_m$.Column(2) then Cancel = True;*
>       display error message "$f_1 \circ \ldots \circ f_n$ must be equal to $g_1 \circ \ldots \circ g_m$!/n" &
>         "Please change accordingly the value(s) of either $f_n$ or/and $g_m$.";
>     *end if;*
>   *end if;*
> *end if;*
> *End Method Form_BeforeUpdate;*

**Figure 3.** *The pseudocode generated by MatBase at the end of the method Form_BeforeUpdate of the class associated to the form built over rdb table dom(g) for enforcing f = g*

Finally, for enforcing the cfdc, *MatBase* generates code in these methods as shown in Figure 4.

Dually, in the (E)MDM an *anti-commutative function diagram constraint* (acfdc) is a function inequality set of the type $f(x) = f_1 \circ \ldots \circ f_n(x) \neq g_1 \circ \ldots \circ g_m(x) = g(x)$, $\forall x \in D$, where $n$ and $m$ are strictly positive naturals and $D$ is the common domain set of both $f$ and $g$. Please note that this is a far stronger constraint than $f \neq g$, which requires only the existence of a $x \in D$ such that $f(x) \neq g(x)$. For example, $(\forall x \in NEIGHBOR\_COUNTRIES)$ $(FrontierColor(Country(x)) \neq FrontierColor(Neighbor(x)))$ is such a constraint, formalizing the rule "for any two neighbor

> *for i = 1 to n, i = i + 1 do*
>   insert at the end of the $f_i\_BeforeUpdate$ method of class $dom(f_i)$ the following pseudocode:
>   "*if not Cancel and not NewRecord and $f_{i-1} \neq f_{i-1}$.OldValue and not IsNull($f_{i-1}$) then*
>   v = *execute*('SELECT $f_1$ FROM $dom(g)$ WHERE $f_n$ IN (SELECT x FROM $dom(f_n)$ …
>       WHERE $f_i$ = ' & x & ')…)' );
>   *if not IsNull(v) then*
>     w = *execute*('SELECT $f_1$ FROM $dom(f_1)$ WHERE x IN (SELECT $f_1$ FROM $dom(f_2)$ …
>         WHERE x = ' & $f_{i-1}$ & ')…)' );
>     *if not IsNull(w) then*

```
    if v ≠ w then Cancel = True;
  display error message 'f_1 ° … ° f_n must be equal to g_1 ° … ° g_m!/n' &
    'You cannot change f_i's value but with one that leaves f_1's unchanged.';
      undo
    end if;
   end if;
  end if;
 end if;"
end if;
end for;
for i = 1 to m, i = i + 1 do
 insert at the end of the g_i_BeforeUpdate method of class dom(g_i) the following pseudocode:
 "if not Cancel and not NewRecord and g_{i-1} ≠ g_{i-1}.OldValue and not IsNull(g_{i-1}) then
  v = execute('SELECT g_1 FROM dom(g) WHERE g_n IN (SELECT x FROM dom(g_m) …
         WHERE g_i = ' & x & ')…)' );
  if not IsNull(v) then
   w = execute('SELECT g_1 FROM dom(g_1) WHERE x IN (SELECT g_1 FROM dom(g_2) …
          WHERE x = ' & g_{i-1} & ')…)' );
   if not IsNull(w) then
    if v ≠ w then Cancel = True;
  display error message 'f_1 ° … ° f_n must be equal to g_1 ° … ° g_m!/n' &
     'You cannot change g_i's value but with one that leaves g_1's unchanged.';
    undo
    end if;
   end if;
  end if;
 end if;"
end for;
```

**Figure 4.** *The* MatBase *algorithm for pseudocode generation at the end of the methods $f_i$ and $g_i$ BeforeUpdate of the classes associated to the forms built over their corresponding rdb tables for enforcing the cfdc $f_1 ° … ° f_n = g_1 ° … ° g_m$*

countries, on any map, the color of their frontiers must be distinct" (where $n = 2$ and $m = 2$).

In the (E)MDM, its case when $n = m = 1$ is also viewed as a hbfp (null-)irreflexivity constraint and its enforcement in *MatBase* is provided in [26].

Moreover, the case when $f$ or $g$ is the unity mapping of a set is called in the (E)MDM a *local anti-commutative fd constraint*. As such fd constraints are of the type $(\forall x \in D)$ $(f(x) \neq x)$, where $f$ is a self-map (autofunction), they are of the type self-map irreflexivity, whose enforcement is dealt with in [24]. For example, *Spouse ° Mother*, *Spouse ° Father*, *Mother ° Spouse*, *Father ° Spouse* are all irreflexive, as parents of somebody may not also be their spouses.

Consequently, in this paper we only deal with the general case of acfdcs, where either $n > 1$ or $m > 1$ and neither $f$ nor $g$ are self-maps. The corresponding combo-boxes are also equipped with the SQL statements presented in Figures 1 and 2. Figure 5 presents the corresponding *Form_BeforeUpdate* method. Finally, for enforcing the acfdc, *MatBase* also generates code as shown in Figure 6.

Enforcing Function Diagram (Anti-)Commutativities

```
Method Form_BeforeUpdate(Boolean Cancel)
Cancel = False
. . .
// enforces constraint f_1 ° … ° f_n(x) ≠ g_1 ° … ° g_m(x), ∀x∈dom(g)
if not Cancel and not IsNull(f_n) and not IsNull(g_m) then
  if not IsNull(f_n.Column(2)) and not IsNull(g_m.Column(2)) then
    if f_n.Column(2) = g_m.Column(2) then Cancel = True;
      display error message "f_1 ° … ° f_n must never be equal to g_1 ° … ° g_m!/n" &
        "Please change accordingly the value(s) of either f_n or/and g_m.";
    end if;
  end if;
end if;
End Method Form_Before Update;
```

**Figure 5.** *The pseudocode generated by* MatBase *at the end of the method Form_Before Update of the class associated to the form built over rdb table dom(g) for enforcing f(x) ≠ g(x), ∀x∈dom(g)*

```
for i = 1 to n, i = i + 1 do
  insert at the end of the f_i_BeforeUpdate method of class dom(f_i) the following pseudocode:
  "if not Cancel and not NewRecord and f_{i-1} ≠ f_{i-1}.OldValue and not IsNull(f_{i-1}) then
  v = execute('SELECT f_1 FROM dom(g) WHERE f_n IN (SELECT x FROM dom(f_n) …
              WHERE f_i = ' & x & ')…)' );
  if not IsNull(v) then
    w = execute('SELECT f_1 FROM dom(f_1) WHERE x IN (SELECT f_1 FROM dom(f_2) …
                WHERE x = ' & f_{i-1} & ')…)' );
    if not IsNull(w) then
      if v = w then Cancel = True;
        display error message ' f_1 ° … ° f_n must never be equal to g_1 ° … ° g_m!/n' &
                  'You cannot change f_i's value but with one that leaves f_1's unchanged.';
        undo
      end if;
    end if;
  end if;
end if;"
end for;
for i = 1 to m, i = i + 1 do
  insert at the end of the g_i_BeforeUpdate method of class dom(g_i) the following pseudocode:
  "if not Cancel and not NewRecord and g_{i-1} ≠ g_{i-1}.OldValue and not IsNull(g_{i-1}) then
  v = execute('SELECT g_1 FROM dom(g) WHERE g_n IN (SELECT x FROM dom(g_m) …
              WHERE g_i = ' & x & ')…)' );
  if not IsNull(v) then
```

```
w = execute('SELECT g₁ FROM dom(g₁) WHERE x IN (SELECT g₁ FROM dom(g₂) …
        WHERE x = ' & g_{i-1} & ')…)' );
if not IsNull(w) then
  if v = w then Cancel = True;
    display error message 'f₁ ° … ° f_n must never be equal to g₁ ° … ° g_m!/n' &
      'You cannot change g_i's value but with one that leaves g₁'s unchanged.';
    undo
  end if;
 end if;
 end if;
end if;"
end for;
```

**Figure 6.** *The MatBase algorithm for pseudocode generation at the end of the methods $f_i$ and $g_i$ BeforeUpdate of the classes associated to the forms built over their corresponding rdb tables for enforcing the acfdc $f_1 \circ \ldots \circ f_n(x) \neq g_1 \circ \ldots \circ g_m(x)$, $\forall x \in dom(g)$*

## RESULTS AND DISCUSSION

For example, according to the algorithms from Figure 1 and 2, *MatBase* generates for the cfdc *Continent = Continent ° Range ° Subrange ° Group ° Mountain* the VBA code presented in Figures 7 and 8, respectively.

Figure 9 shows the VBA code generated according to the algorithm in Figure 3 and then manually enhanced with a friendly error message, as was the case for all subsequent methods.

```
SELECT MOUNTAINS.x, [MOUNTAIN_RANGES].[Range] & ", " &
       [MOUNT_SUBRANGES].[Subrange] & ", " & [MountGroup] & ", " &
       [Mountain] AS [Range, Subrange, Group, Mountain],
       MOUNTAIN_RANGES.Continent
       FROM MOUNTAIN_RANGES RIGHT JOIN
       (MOUNT_SUBRANGES RIGHT JOIN (MOUNT_GROUPS RIGHT JOIN
       MOUNTAINS ON MOUNTAINS.Group = MOUNT_GROUPS.x) ON
       MOUNT_SUBRANGES.x = MOUNT_GROUPS.Subrange) ON
       MOUNTAIN_RANGES.x = MOUNT_SUBRANGES.Range
ORDER BY [MOUNTAIN_RANGES].[Range] & ", " &
       [MOUNT_SUBRANGES].[Subrange] & ", " & [MountGroup] & ", " &
       [Mountain];
```

**Figure 7.** *The SQL statement generated by MatBase as the row source for the combo-box Mountain of the form RIVERS*

```
SELECT [CONTINENTS].[x], [CONTINENTS].[Continent] FROM CONTINENTS
ORDER BY [Continent];
```

**Figure 8**. *The SQL statement generated by MatBase as the row source for the combo-box Continent of the form RIVERS (i.e., same as for MOUNTAIN_RANGES)*

Because $m = 1$, as shown in Figure 8, there is no need for a third column in that combo-box. However, because $n > 1$, as shown in Figure 7, *Mountain.Column*(2) is a string, so *Continent* from *RIVERS*, which is a foreign key, so an integer, had to be converted to a string (using the VBA *CStr* function).

To also alert users audibly when errors occur, *MatBase* is also generating a *Beep* statement.

Figures 10 to 13 present the code generated by *MatBase* for enforcing this cfdc according to the algorithm from Figure 4. Note that, as *Group* is a reserved VBA (and SQL) keyword, the name of the homonym column must be embraced in square parentheses.

```
Sub Form_BeforeUpdate(Cancel As Integer)
. . .
'enforces constraint Continent = Continent ° Range ° Subrange ° Group ° Mountain
If Not Cancel And Not IsNull(Mountain) Then
  If Not IsNull(Mountain.Column(2)) Then
    If Mountain.Column(2) <> CStr(Continent) Then
      Cancel = True
      Beep
      MsgBox "The mountain from which a river springs must be on the same " & _
        "continent as the river!" & Chr(13) & "The " & River & " flows through " & _
        DLookup("Continent", "CONTINENTS", "x =" & Continent) & _
          ", while " & Mountain.Column(1) & " is in " & DLookup( _
          "Continent", "CONTINENTS", "x =" & Mountain.Column(2)) & _
          ": please change either the continent or/and the mountain!", _
        vbCritical, "Request rejected... "
    End If
  End If
End If
End Sub
```

**Figure 9.** *The VBA code generated by MatBase at the end of the Form_BeforeUpdate method of the class associated to the form RIVERS*

According to the algorithm from Figure 2, *MatBase* generates the code from Figure 8 for both *Country* and *Neighbor* defined on *NEIGHBOR_COUNTRIES* to enforce the acfdc (∀x∈NEIGHBOR_COUNTRIES)(*FrontierColor*(*Country*(x)) ≠ *FrontierColor*(*Neighbor*(x)).

Figure 14 shows the code generated by *MatBase* according to the algorithm from Figure 5. Finally, Figure 15 shows the one generated according to the algorithm from Figure 6.

Please note that, as null values do not violate either commutative or anti-commutative constraints, they are ignored everywhere in the generated code, which makes it run the fastest possible.

```
Sub Continent_BeforeUpdate(Cancel As Integer)
…
'enforces constraint Continent = Continent ° Range ° Subrange ° Group ° Mountain
Dim v As Variant
If Not Cancel And Not NewRecord And Continent <> Continent.OldValue Then
  v = DLookup("River", "RIVERS", "Continent =" & Continent.OldValue & _
    " AND Mountain IN (SELECT x FROM MOUNTAINS WHERE [GROUP] IN " & _
    "(SELECT x FROM MOUNT_GROUPS WHERE Subrange IN " & _
    "(SELECT x FROM MOUNT_SUBRANGES WHERE Range =" & x & ")))")
  If Not IsNull(v) Then
    Cancel = True
```

```
      Beep
      MsgBox "The Mountain from which a river springs must be on the same continent as " & _
         "the river!" & Chr(13) & "At least the river " & v & ", which flows through " _
         & DLookup("Continent", "CONTINENTS", "x =" & Continent.OldValue) _
         & ", springs from a Mountain belonging to this mountain range:" _
         & Chr(13) & "you cannot change this continent!", vbCritical, "Request rejected..."
      Undo
    End If
  End If
End Sub
```

**Figure 10.** *The VBA code generated by MatBase at the end of the Continent_BeforeUpdate method of the class associated to the form MOUNTAIN_RANGES*

## CONCLUSION

We analyzed enforcement of commutative and anti-commutative function diagram constraints in db software applications. We proved that the best corresponding architectural approach is using event-driven, SQL embedding high level programming languages.

```
Sub Range_BeforeUpdate(Cancel As Integer)
…
'enforces constraint Continent = Continent ° Range ° Subrange ° Group ° Mountain
Dim v, w As Variant
If Not Cancel And Not NewRecord And Range <> Range.OldValue And _
Not IsNull(Range) Then
  v = DLookup("Continent", "Rivers", "Mountain IN (SELECT x FROM MOUNTAINS " & _
        "WHERE [Group] IN (SELECT x FROM MOUNT_GROUPS " & _
        "WHERE Subrange =" & x & "))")
  If Not IsNull(v) Then
    w = DLookup("Continent", "MOUNTAIN_RANGES", "x =" & Range)
    If Not IsNull(w) Then
      If CLng(v) <> CLng(w) Then
        Cancel = True
        Beep
        MsgBox "The Mountain from which a river springs must be on the same continent as " _
            & "the river!" & Chr(13) & "At least the river " & DLookup("River", "RIVERS", _
            "Mountain =" & x) & ", which flows through " & DLookup("Continent", _
            "CONTINENTS", "x = DLookup ('Continent', 'RIVERS', 'Mountain =' & x)") & _
            ", springs from a Mountain of a group of this subrange:" & Chr(13) & _
            "you cannot change this mountain range but with one from this continent!", _
            vbCritical, "Request rejected..."
        Undo
      End If
    End If
  End If
End If
End Sub
```

**Figure 11.** *The VBA code generated by MatBase at the end of the Range_BeforeUpdate method of the class associated to the form MOUNT_SUBRANGES*

```
Sub Subrange_BeforeUpdate(Cancel As Integer)
…
'enforces constraint Continent = Continent ° Range ° Subrange ° Group ° Mountain
Dim v, w As Variant
If Not Cancel And Not NewRecord And Subrange <> Subrange.OldValue And _
Not IsNull(Subrange) Then
  v = DLookup("Continent", "Rivers", "Mountain IN (SELECT x FROM MOUNTAINS " _
        & "WHERE [Group] =" & x & ")")
  If Not IsNull(v) Then
        w = DLookup("Continent", "MOUNTAIN_RANGES", " x IN (SELECT Range " & _
       "FROM MOUNT_SUBRANGES WHERE x =" & Subrange & ")")
    If Not IsNull(w) Then
      If CLng(v) <> CLng(w) Then
        Cancel = True
        Beep
        MsgBox "The Mountain from which a river springs must be on the same continent as " _
        the river!" & Chr(13) & "At least the river " & DLookup("River", "RIVERS", _
           "Mountain =" & x) & ", which flows through " & DLookup("Continent",
           "CONTINENTS", "x = DLookup ('Continent', 'RIVERS', 'Mountain =' & x)") & _
           ", springs from a Mountain of this group:" & Chr(13) & _
           "you cannot change this mountain subrange but with one from this continent!", _
           vbCritical, "Request rejected..."
        Undo
      End If
    End If
  End If
End If
End Sub
```

**Figure 12.** *The VBA code generated by MatBase at the end of the Subrange_BeforeUpdate method of the class associated to the form MOUNT_GROUPS*

```
           Sub Group_BeforeUpdate(Cancel As Integer)
…
'enforces constraint Continent = Continent ° Range ° Subrange ° Group ° Mountain
Dim v, w As Variant
If Not Cancel And Not NewRecord And Group <> Group.OldValue And Not IsNull(Group) _
Then
  v = DLookup("Continent", "Rivers", "Mountain =" & x)
  If Not IsNull(v) Then
    w = DLookup("Continent", "MOUNTAIN_RANGES", " x IN (SELECT Range " & _
Sub Form_BeforeUpdate(Cancel As Integer)
```

Enforcing Function Diagram (Anti-)Commutativities

```
            "FROM MOUNT_SUBRANGES WHERE x IN (SELECT Subrange FROM " & _
            "MOUNT_GROUPS WHERE x =" & Group & "))")
    If Not IsNull(w) Then
      If CLng(v) <> CLng(w) Then
        Cancel = True
        Beep
        MsgBox "The Mountain from which a river springs must be on the same continent as " _
            "the river!" & Chr(13) & "At least the river " & DLookup("River", "RIVERS", _
            "Mountain =" & x) & ", which flows through " & DLookup("Continent", _
            "CONTINENTS", "x = DLookup ('Continent', 'RIVERS', 'Mountain =' & x)") & _
            ", springs from this Mountain:" & Chr(13) & "you cannot change this group!", _
            vbCritical, "Request rejected..."
        Undo
      End If
    End If
  End If
End If
End Sub
```

**Figure 13.** *The VBA code generated by MatBase at the end of the Group_BeforeUpdate method of the class associated to the form MOUNTAINS*

```
Sub Form_BeforeUpdate(Cancel As Integer)
…
'enforces the constraint FrontierColor(Country(x)) <> FrontierColor(Neighbor(x))
Dim v, w As Variant
v = DLookup("FrontierColor", "COUNTRIES", "x =" & Country)
If Not IsNull(v) Then
  w = DLookup("FrontierColor", "COUNTRIES", "x =" & Neighbor)
  If Not IsNull(w) Then
    If v = w Then
      Cancel = True
      Beep
      MsgBox "Frontier colors of neighbor countries must be distinct!" & Chr(13) & _
          "For both " & DLookup("Country", "COUNTRIES", "x =" & Country) & " and " & _
          DLookup("Country", "COUNTRIES", "x =" &  Neighbor) & ", color " & v & _
          " is used:" & Chr(13) & "please either change Country or/and Neighbor or/and the " _
          & "frontier color of at least one of them.", vbCritical, "Request rejected..."
    End If
  End If
End If
End Sub
```

**Figure 14.** *The VBA code generated by MatBase at the end of the Form_BeforeUpdate method of the class associated to the form NEIGHBOR_COUNTRIES*

We presented the pseudocode algorithms used by *MatBase*, an intelligent DBMS prototype, to generate code for enforcing these two dual types of constraints. Of course, these algorithms may also be used manually by developers not having access to a *MatBase* copy.

We stressed that not enforcing all business rules in db schemes and apps eventually results in storing unplausible data in dbs, which, in its turn, leads to poor quality information and knowledge extracted from them through queries and reports.

We then presented and discussed the VBA code generated by *MatBase* for enforcing one commutativity and one anti-commutativity constraints from a Geography app.

```
Sub FrontierColor_BeforeUpdate(Cancel As Integer)
…
'enforces the constraint FrontierColor(Country(x)) <> FrontierColor(Neighbor(x))
Dim v As Variant
If Not IsNull(FrontierColor) Then
  v = DLookup("Country", "NEIGHBORS", "Neighbor =" & x & _
              " AND Country IN (SELECT x FROM COUNTRIES " & _
              "WHERE FrontierColor =" & FrontierColor & ")")
  If Not IsNull(v) Then
    Cancel = True
  Else
    v = DLookup("Neighbor", "NEIGHBORS", "Country =" & x & _
                " AND Neighbor IN (SELECT x FROM COUNTRIES " & _
                "WHERE FrontierColor =" & FrontierColor & ")")
    If Not IsNull(v) Then Cancel = True
  End If
End If
If Cancel Then
 Beep
  MsgBox "Frontier colors of neighbor countries must be distinct!" & Chr(13) & Country & _
        " is neighbor to " & DLookup("Country", "COUNTRIES", "x =" & v) & _
        ", which also has " & FrontierColor.Column(1) & " as frontier color:" & Chr(13) & _
        "Please change frontier color.", vbCritical, "Request rejected..."
 Undo
End If
End Sub
```

**Figure 15.** *The VBA code generated by MatBase at the end of the FrontierColor_BeforeUpdate method of the class associated to the form COUNTRIES*

This code was then enhanced by replacing its syntactical-type error messages with more friendly semantical ones.

The approach we advocate in this paper is a new step towards replacing the current ad-hoc db software development strategy with the paradigm of modeling as programming [18, 34, 35].

**Conflict of Interest**

The authors declare that the research was conducted in the absence of any commercial or financial relationships that could be construed as a potential conflict of interest.

**Author Contributions**

Christian Mancas wrote the first two sections of this paper.

Diana Christina Mancas wrote the last two ones, as well as the corresponding *MatBase* code, both in VBA and C#.

Enforcing Function Diagram (Anti-)Commutativities31. Dyer L., et al. (2012). *Scaling BPM Adoption from Project to Program with IBM Business Process Manager*. http://www.redbooks.ibm.com/redbooks/pdfs/sg247973.pdf.

32. Red Hat Customer Content Services. (2024). Getting Started with Red Hat Business Optimizer. https://access.redhat.com/documentation/en-us/red_hat_decision_manager/7.1/html/getting_started_with_red_hat_business_optimizer/index.

33. Agiloft Inc. (2022). *Agiloft Reference Manual*, https://www.agiloft.com/documentation/agiloft-developer-guide.pdf.

34. Thalheim B. (2020). Models as programs. In: Bork D., Karagiannis D., Mayr H.C. (eds.): Modellierung 2020, Lecture Notes in Informatics (LNI), Gesellschaft für Informatik, Bonn, Germany, pp. 193–195.

35. Mancas C. (2020). On Modelware as the 5th Generation of Programming Languages. ASCS 2(9):24–26.
Open Access Journal of Computer Science and Engineering V1. I1. 2024     13